\documentclass[10pt, conference, letterpaper]{IEEEtran}
\IEEEoverridecommandlockouts
\usepackage{cite}
\usepackage{amsmath,amssymb,amsfonts}
\usepackage{graphicx}
\usepackage{textcomp}
\usepackage{xcolor}
\usepackage{subcaption}
\usepackage{algpseudocode}
\usepackage[linesnumbered,lined,ruled]{algorithm2e}
\usepackage{multirow}
\usepackage{array}
\usepackage{makecell}
\usepackage{caption}
\usepackage{tikz}
\usepackage{graphicx}
\usepackage{subcaption}
\usepackage{booktabs}
\usepackage{ragged2e} 
\usepackage{hyperref}
\usepackage{cleveref}

\SetCommentSty{mycommfont}
\SetKwComment{Comment}{// }{}
\usepackage{inconsolata}
\usepackage{tcolorbox}
\newtheorem{insight}{Insight}
\def\BibTeX{{\rm B\kern-.05em{\sc i\kern-.025em b}\kern-.08em
    T\kern-.1667em\lower.7ex\hbox{E}\kern-.125emX}}
\begin{document}

\title{HALO: Semantic-Aware Distributed LLM Inference \\in Lossy Edge Network}

\author{\IEEEauthorblockN{Peirong Zheng$^{1}$, Wenchao Xu*$^{2}$, Haozhao Wang$^{3}$, Jinyu Chen$^{1}$, and Xuemin (Sherman) Shen$^{4}$,~\IEEEmembership{Fellow, IEEE}}
\IEEEauthorblockA{
\textsuperscript{1}Department of Computing, The Hong Kong Polytechnic University\\
\textsuperscript{2}Division of Integrative Systems and Design, The Hong Kong University of Science and Technology\\
\textsuperscript{3}School of Computer Science and Technology, Huazhong University of Science and Technology\\
\textsuperscript{4}Department of Electrical and Computer Engineering, University of Waterloo \\
peirong.zheng@connect.polyu.hk, wenchaoxu@ust.hk, hz\_wang@hust.edu.cn \\
jinyu.chen@connect.polyu.hk, sshen@uwaterloo.ca\\
*Corresponding author: Wenchao Xu}
}

\maketitle
\newcommand{\Name}{\texttt{HALO}\space}
\newcommand{\name}{\texttt{HALO}} %
\begin{abstract}

The deployment of large language models' (LLMs) inference at the edge can facilitate prompt service responsiveness while protecting user privacy. However, it is critically challenged by the resource constraints of a single edge node. Distributed inference has emerged to aggregate and leverage computational resources across multiple devices. Yet, existing methods typically require strict synchronization, which is often infeasible due to the unreliable network conditions. In this paper, we propose \Name, a novel framework that can boost the distributed LLM inference in lossy edge network. The core idea is to enable a relaxed yet effective synchronization by strategically allocating less critical neuron groups to unstable devices, thus avoiding the excessive waiting time incurred by delayed packets. \Name introduces three key mechanisms: (1) a semantic-aware predictor to assess the significance of neuron groups prior to activation.
(2) a parallel execution scheme of neuron group loading during the model inference.
(3) a load-balancing scheduler that efficiently orchestrates multiple devices with heterogeneous resources.
Experimental results from a Raspberry Pi cluster demonstrate that HALO achieves a 3.41x end-to-end speedup for  LLaMA-series LLMs under unreliable network conditions. It maintains performance comparable to optimal conditions and significantly outperforms the state-of-the-art in various scenarios.

\end{abstract}


\section{Introduction}

Recently, massive natural language processing (NLP) applications, such as voice assistants and chatbots, machine translation, etc., are emerging in edge environments with a focus on preserving the privacy of users and providing prompt responses\cite{king2024sasha,li2024personal}. To implement these applications, various transformer models are deployed on edge devices. However, as transformer models, particularly large language models (LLMs), continue to grow in scale, their deployment has become increasingly challenging due to the inherent computational and memory constraints of edge devices.

\begin{figure}[t]
    \centering
    \includegraphics[width=0.73\linewidth]{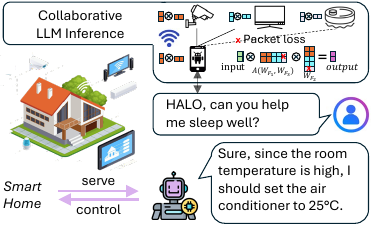}
    \caption{Smart home scenario empowered by \name: trusted home devices collaboratively serve LLM inference with high accuracy and efficiency in lossy network.}
    \label{fig:scenario}
    \vspace{-4pt}
\end{figure}

To tackle this challenge, many approaches propose deploying LLMs on multiple edge devices in a distributed manner via parallel computing. For instance, trusted edge devices in smart home scenario can contribute their computing resources for collaborative LLM inference within a local area network \cite{ye2024galaxy}\cite{xu2024deploying}, as shown in \Cref{fig:scenario}.
Existing methods include intra-layer tensor parallelism (TP) \cite{shoeybi2019megatron}, inter-layer pipeline parallelism (PP)\cite{zhao2023lingualinked}, and sequence parallelism (SP)\cite{korthikanti2023reducing}. For instance, to deal with the resource heterogeneity of multiple edge devices, \textit{LinguaLinked} employs inter-layer PP to increase throughput by considering both the constraints of each device and the inter-dependencies between the layers of the LLMs, which requires accurate synchronization among multiple nodes for an overlapping scheme \cite{zhao2023lingualinked}.
Further, considering the one-shot (referring to a single request) inference characteristic of LLM applications in edge, intra-layer TP is gaining broader adoption to reduce the inference latency. 
For example, \textit{EdgeShard} combines intra-layer TP and inter-layer PP to optimize latency and throughput by considering the heterogeneities in both edge node and servers \cite{zhang2024edgeshard}. 
\textit{Galaxy} combines SP and intra-layer TP, and employs a tile-wise overlapping method to parallel the communication and model execution processes \cite{ye2024galaxy}. 

While existing distributed inference mechanisms have achieved considerable success, yet they generally presuppose a stable network environment—an assumption that is rarely feasible in most edge computing scenarios \cite{bhadra2015packet,ali2024congestion}.
For instance, the end-to-end packet loss rate (PLR) can be up to $5\%$ even under in-door conditions \cite{sheshadri2017packet}. For cellar networks, the PLR can be up to $10\%$ and even worse in poor channel areas \cite{9524600}. The synchronization overhead for distributed inference can be greatly increased due to a relatively high PLR, and thus compromise the LLM performance, e.g, 4.26x slower for the time per output token (TPOT) under 5\% PLR in our \hyperref[tab-communication-time]{preliminary experiment}.

To enable efficient distributed LLM inference in lossy edge network, we propose \name, a semantic-aware algorithm-system co-design to aggregate and leverage computational resources from multiple edge devices with unreliable network connections.  
The core idea of \Name is to allocate less critical neuron groups of LLM to unreliable edge devices while facilitating relaxed yet effective synchronization. \Name enables the inference process to continue without waiting for overly delayed packets while maintaining the inference accuracy. Specifically, the key innovations of \Name are:
\vspace{-3pt}
\begin{itemize}
    \item Runtime semantic-aware predictors that identify neuron groups importance for mapping and loading, hence ensuring the effectiveness even for relaxed synchronization.
    \item An overlapping scheme that parallel the neuron group mapping \& loading process with communication and computation, to overcome the system bottleneck and accelerate the LLM inference.  
    \item  A PLR-aware scheduler that efficiently adapts to the edge heterogeneity in memory budget and computing power for load balancing and straggler minimization.
\end{itemize}
To the best of our knowledge, \Name represents the first comprehensive framework to accelerate LLM inference specifically designed for lossy edge network. 
We address the challenges of slow collaborative inference under packet loss between heterogeneous edge devices by proposing semantic-aware methods to achieve effective relaxed synchronization, combined with meticulous overlapping scheme and PLR-aware load-balancing scheduler to maximize resource efficiency.  
We implement \Name on a real-world testbed, composed of 8 typical edge devices, Raspberry Pi 4 Model B (RPi), with LLaMA-series LLMs. It achieves 3.41x end-to-end speedup under 5\% PLR while preserving comparable accuracy, demonstrating significant improvements compared to existing approaches.

\section{Background and Motivation}
\label{sec:background}
\subsection{LLM Inference over Edge Devices}
\noindent \textbf{Transformer decoder-based LLMs.}
Modern LLMs follow the generative pretrained transformer (GPT) paradigm, composed of the token embedding table, multiple Transformer decoder layers, and the language modeling head (LM Head). The token embedding table resides solely in memory without requiring computational resources. In the Transformer decoder layer, the multi-head attention (MHA) and multi-layer perceptron (MLP) blocks are the primary computational and memory components suitable for distribution. In addition, the LM Head is also a distributable linear layer. The remainder is element-wise operations connecting these layers, including dropout, normalization, and residual connection.

\noindent\textbf{Distributed LLM inference over edge.}
In the edge scenario, the one-shot inference, such as the chatbot, is dominantly popular. Intra-layer TP\cite{shoeybi2019megatron} is designed for LLM training in the data center, and evenly assigns the workload to all GPUs. It is suitable for reducing latency in one-shot inference \cite{ye2024galaxy,li2024tpi,dllama}. 
Hence, we propose TP-based methods to utilize each device's computing power and memory simultaneously. 
TP is composed of consecutive block-wise computation and activation synchronization. The algorithm performance is heavily determined by the straggler (i.e., the one that finishes work last), which can leave other devices idle and increase the overall inference latency. Hence, load balance is of vital importance. Second, the synchronization overhead becomes the bottleneck due to unstable links according to the preliminary experiment on the relation of inference time and PLR.

\Cref{tab-communication-time} shows the relation between different PLRs (0-5\%) and TPOT to validate the influence of unstable links and efficiency of unreliable transmission. Eight RPis are involved and benchmarked by TP framework \texttt{dllama}\cite{dllama} and model TinyLlama\cite{zhang2024tinyllama}. TCP is the default communication protocol providing reliable transmission. UDP with timeout is the protocol for efficiency and provides unreliable transmission.
A timeout mechanism is essential because, in its absence, a receiver would wait indefinitely for lost packets, thereby stalling the entire inference process.
Hence, we limit the waiting time within the timeout of 10ms empirically. For TCP, as the PLR increases by 1\%, the TPOT can increase by about 617.2ms. When the PLR=1\%, its overhead nearly accounts for 70\% of the TPOT without PLR. Obviously, current communication optimization approaches, such as overlap, will lose effectiveness and cause downgraded system performance in a lossy network.
For UDP, as the PLR increases by 1\%, the TPOT increases slightly by about 28.4ms. It achieves at most 4.26x speedup than TCP under 5\% PLR. 
Using UDP for uploading packets also allows for faster synchronization, without the overhead of TCP mechanisms under 0\% PLR.
However, simply adopting UDP transmission could suffer from packet loss. From the application view, model accuracy is severely degraded by activation loss (encapsulated in packets). To preserve model accuracy and exploit the efficiency brought by sacrificed transmission reliability, we need information about packets' importance prior to upload and reserve the transmission reliability for important packets.

\begin{table}[t]
\centering
\caption{Comparison of Average TPOT of 64 tokens.}
\label{tab-communication-time}
\begin{tabular}{l cc}
\toprule
\multirow{2}{*}{PLR (\%)} & \multicolumn{2}{c}{ TPOT (ms)} \\
\cmidrule(l){2-3} %
& TCP   & UDP / 10ms timeout  \\ 
\midrule 
0 &899.45 &793.10 \\
1 &1531.53 &765.66 \\
2 &2068.10 &818.48 \\
5 &3983.52 &934.95 \\
\bottomrule %
\end{tabular}
\end{table}

\noindent\textbf{Neuron group.}
TP distributes model weights to each device, with dependent neurons grouped in one device. In the MHA, one head and its output projection weight are an independent neuron group, while in the MLP, each up-projection, gate function, and down-projection can be viewed as one independent neuron group; the LM Head only contains up-projection, which is also a collection of neuron groups. For hardware-friendly implementation and compatibility with quantization, we fix the group size to 256. For instance, the MLP with dimension 6,144 can be divided into 24 distributable neuron groups. We adopt the term neuron groups to refer to the dependent neurons in one block (MHA/MLP/LM Head).

\subsection{Insight: Semantic-Aware Synchronization}
\label{insight1}
\begin{tcolorbox}
\begin{insight}
Semantic imbalance (activation sparsity) enables relaxed synchronization to be effective.
\end{insight}
\end{tcolorbox}

\begin{figure}[t]
    \centering
    \begin{subfigure}{0.48\columnwidth}
        \centering
        \includegraphics[width=\columnwidth, trim=0 0 0 2cm, clip]{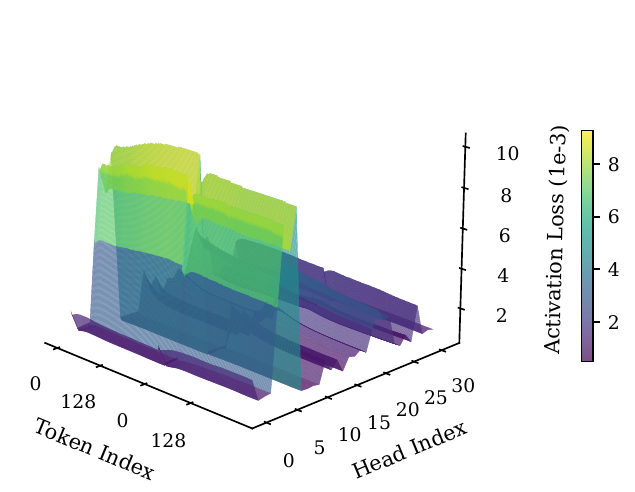}
        \caption{Activation sparsity of neuron groups in the MHA block.}
    \end{subfigure}
    \hfill
    \begin{subfigure}{0.48\columnwidth}
        \centering
        \includegraphics[width=\columnwidth, trim=0 0 0 2cm, clip]{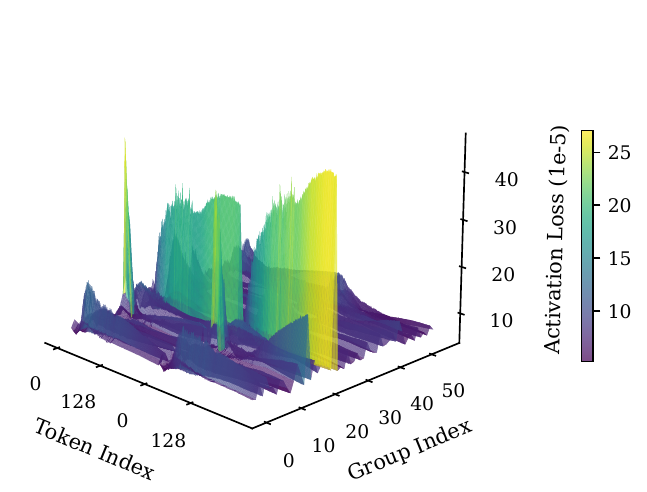}
        \caption{Activation sparsity of neuron groups in the MLP block.}
    \end{subfigure}
    \caption{Activation Distribution of Neuron Groups.}
    \label{fig:sparsity}
\end{figure}

To optimize activation synchronization of TP in lossy edge network, we delve into the activation distribution during LLM inference.
\Cref{fig:sparsity} presents the skewed activation distribution among neuron groups. 
The semantic features are the merging result of all neuron groups' activation. The input is two random samples from GSM8K\cite{cobbe2021gsm8k}, and the output is layer 3 of Llama-3.1-8B-Instruct-int4 model. 
The loss of activations refers to the L2 norm of an activation, which indicates its importance.
Both (a) and (b) demonstrate that activation patterns exhibit sparsity, where certain heads/groups dominate the merged semantic features at the token level while others maintain minimal contribution. The semantic features, are largely determined by top activations, a subset of total activations.
Besides, the two samples' activation distribution patterns are different, especially in the MLP block. This dynamic characteristic limits the effectiveness of neuron group importance statistics. 

\setlength{\textfloatsep}{5pt}  %
We conduct a small-scale experiment to verify the imbalanced impact of the neuron groups' activation. In \Cref{tab:activation_loss}, all layers exhibit the same activation loss method, under 10\% PLR and 25\% of neuron groups. The perturbed semantic features, which are influenced by the activation loss, change the inference accuracy.
It shows that loss of high-norm activation causes a significant accuracy degradation of 47.2\%, while loss of low-norm activation degrades merely 1.6\% accuracy. The performance of random activation loss is roughly in the middle of the other two methods. This proves that a subset of activations can still preserve comparable inference accuracy.
The low-norm activation loss method is not realistically useful because this importance can only be known after the neuron groups have already been computed.
Given that semantic features change slowly across layers \cite{liu2023deja}, it is possible to train highly-accurate neural networks to learn the output activation distribution using input semantic features from layers ahead of the current layer.

\begin{table}[t]
\centering
\caption{TinyGSM8K\cite{polo2024tinybenchmarks} Accuracy of Llama-3.1-8B-Instruct-int4 with Different Activation Loss Methods.}
\label{tab:activation_loss}
\begin{tabular}{lcr}
\toprule
\textbf{Method} & \textbf{Accuracy (\%)} & \textbf{$\Delta$ (\%)} \\
\midrule
Baseline (No Loss) & 71.6 & - \\
Random activation loss & 44.4 & -27.2 \\
High-norm activation loss & 24.4 & -47.2 \\
Low-norm activation loss & 70.0 & -1.6 \\
\bottomrule
\end{tabular}
\end{table}

\noindent\textbf{Opportunities of semantic-aware synchronization.}
In a lossy network, with SAPs, we could discard strict slow synchronization and enjoy the efficiency of relaxed yet effective synchronization, as shown in the \Cref{tab-communication-time}.
The benefits are: 1) It reduces synchronization overhead under lossy network; 2) It preserves critical activations, thus maintaining accuracy.

\begin{figure}[t]
    \centering
    \includegraphics[width=\linewidth]{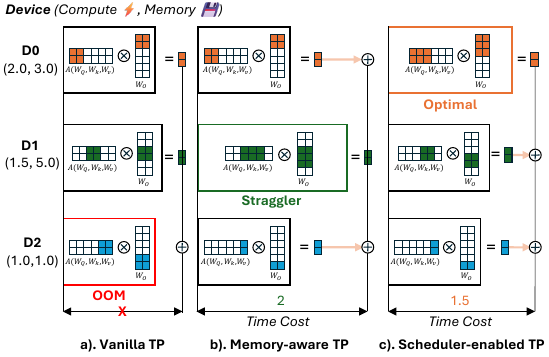}
    \caption{Time Cost of Different Scheduling Strategies.}
    \label{fig:scheduler}
\end{figure}
\subsection{Insight: Comprehensive Heterogeneity-aware Scheduler}

\begin{tcolorbox}
    \begin{insight}
        PLR-awareness is fully compatible with load balancing algorithms. The scheduler determines the runtime mapping priority to optimize communication, while load balancing algorithms optimize computation.
    \end{insight}
\end{tcolorbox}

Edge environments are characterized by profound heterogeneity in both memory budgets and computing power, posing significant challenges for LLM inference. To tackle these challenges, we start by delving into the characteristics of LLMs' layers, essentially matrix parameters, which contribute linearly to both memory and computing requirements. This attribute makes it possible to allocate the workload linearly by the number of neuron groups. We use one attention block as an example to illustrate the critical need for a comprehensive scheduler in \Cref{fig:scheduler}. The attention block is distributed along the head dimension to three devices. The number of filled squares represents the workload, and the vertical line with addition represents the merging operation. A naive approach like Vanilla TP (a), which ignores resource limits, immediately fails with an out-of-memory (OOM) error on the memory-constrained device D2. An improved Memory-aware TP (b) avoids OOM but creates a new performance bottleneck by overloading the memory-sufficient \& compute-slow device D1, turning it into a straggler that dictates the overall latency while other devices sit idle. This highlights that heuristic optimization for memory and compute is suboptimal. Scheduler-enabled TP (c) demonstrates the power of a holistic approach. By jointly considering both memory budgets and computing power, it achieves load balancing, mitigates straggler effect, and minimizes inference time. This insight drives us to utilize all available resources for optimal performance.

\begin{figure*}[ht]
    \centering
    \includegraphics[width=0.95\textwidth]{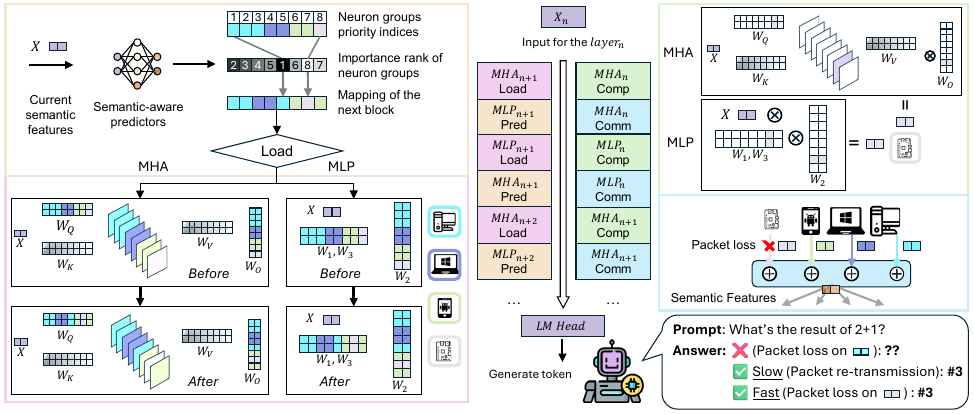}
    \caption{\textit{Generation} Phase with Overlapped Processes. Load\&Comp and Pred\&Comm are overlapped across layers during runtime. Four processes have four colors, and the block lines, containing detailed illustrations, are in the same color. For illustration simplicity, we ignore the LM Head, which is essentially similar to the MLP block.}
    \label{fig:predictor}
\end{figure*}

As shown in \Cref{tab-communication-time}, packet loss can severely increase the communication time. To reduce this latency, an intuitive method is to alter the load distribution to evict devices associated with high PLR. The rationale would be to reduce communication overhead by reducing the participating devices. However, such a strategy decreases the total computing power, increases computation time, and introduces the need for a complex trade-off optimization. The presence of PLR needs an informative scheduler that preserves critical PLR information without interfering with load balance. Accordingly, our scheduler is designed to integrate the PLR order of each participant directly into its schedule, while strictly upholding the optimal load balance. The core of our approach is a simple yet effective technique: we transform neuron group indices into \textit{neuron group priority indices} (see details in ~\Cref{alg:scheduler}). These priority indices, in conjunction with real-time results from our SAPs, guide the dynamic updates of neuron groups (see Pred process in \Cref{fig:predictor}). This mechanism enables us to obtain the communication time savings afforded by relaxed synchronization, without incurring increased computation time.

\noindent\textbf{Opportunities of comprehensive scheduling.}
A comprehensive scheduler offers a dual advantage: 1). It is designed to guarantee optimal load balance across heterogeneous devices. 2). It embeds PLR-awareness directly into the scheduling indices, enabling highly efficient, dynamic updates at runtime.

\setlength{\dbltextfloatsep}{0pt}

\begin{figure*}[]
    \centering
    \includegraphics[width=0.95\textwidth]{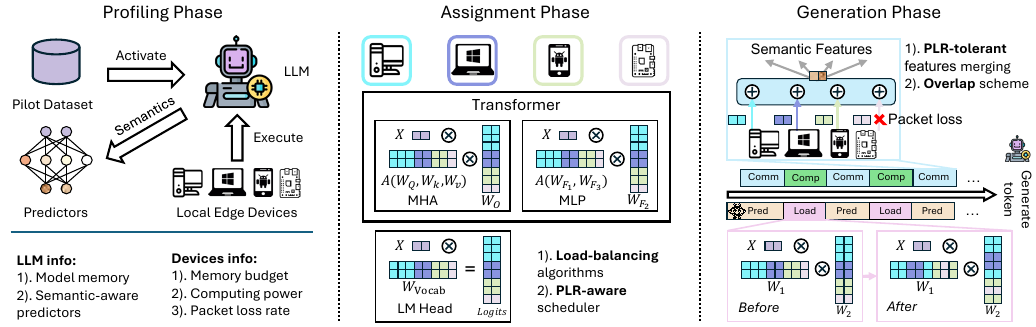}
    \caption{Overview of the System Design. Three phases are executed in sequence. Four edge devices collaboratively inference one LLM. In the \textit{Assignment} phase, their four colors and the number of filled squares indicate the assigned workload. $A()$ refers to the attention computation in the MHA block and activation function computation in the MLP block.}
    \label{fig:system_overview}
\end{figure*}

\section{System Design}
\subsection{Overview}
\Cref{fig:system_overview} presents \name, which features three phases: \textit{Profiling Phase, Assignment Phase, and Generation Phase}. \textit{Profiling} phase profiles both the model and edge devices. \Name profiler utilizes a prior dataset to gather the activation of neuron groups and train the SAPs accordingly, which is an offline one-time operation. We exploit the temporal stability of edge environments by conducting pre-inference profiling, assuming conditions remain almost constant during execution. Before each inference, \Name profiler receives user-assigned memory budgets and utilizes Linux tools to obtain the devices' information, including computing power and PLR. This profiling information is then used in the following phases.

\textit{Assignment} phase is the generation of the scheduler for each device. It aims to map devices with neuron group priority indices. Based on profiling the model and devices, the scheduler can determine the workload ratio and then map the neuron group priority indices to each device to ensure load balance. The indices are a list of integer sequence numbers.

\textit{Generation} starts with the master node's command. Then, each worker reads the assigned neuron groups according to the scheduler. After the full model has been loaded, the generation process begins. During the Generation, Computation (Comp) and communication (Comm) of TP are executed in the system until all layers have been forwarded. The predicting (Pred) and loading (Load) processes are executed in parallel with the original TP inference. Finally, the master node gathers the logits for token decoding and continues this auto-regressive process to generate the next token. 
This parallel execution scheme is detailed in \Cref{fig:predictor} and elaborated in \Cref{subsec:plr-generation}.

\subsection{Workflow and Algorithms} 
We consider 3 aspects: memory budget, computing power, and PLR. The top priority is the total memory budget, which determines whether the model and KV cache can be accommodated. \Cref{tab:notations} includes the notation for model and device informations. We denote the workload for the MHA, MLP, and LM Head in the same form of neuron groups in the algorithms.

\begin{table}[t]
\centering
\caption{Notations for Formulas and Algorithms}
\label{tab:notations}
\begin{tabular}
{>{\RaggedRight}p{1.2cm}p{6.5cm}}
\toprule
\textbf{Notation} & \textbf{Definition} \\
\midrule
\multicolumn{2}{l}{\textit{\textbf{Device-Related Notations}}} \\
$n$ & The total number of available devices. \\
$i$ & The index for a device, where $i \in \{1, \ldots, n\}$. \\
$m_i$ & The memory budget (e.g., in MB) of device $i$. \\
$c_i$ & The computing power (e.g., in TFLOPS) of device $i$. \\
$\boldsymbol{r}$ & The vector of memory allocation ratios for $n$ devices, $\boldsymbol{r} = (r_1, \ldots, r_n)$, where $\sum_{i=1}^{n} r_i = 1$. \\
$\boldsymbol{\rho}$ & The vector of PLRs for $n$ devices, $\boldsymbol{\rho} = (\rho_1, \ldots, \rho_n)$. \\
\midrule
\multicolumn{2}{l}{\textit{\textbf{Model-Related Notations}}} \\
$\mathcal{M}_{t}$ & The total memory demand of the model. \\
$L$ & Number of Transformer decoder layers in the model. \\
$D$ & The dimension of the hidden states. \\
$N_h$ & Number of attention heads in one MHA block. \\
$N_{kv}$ & Number of key-value (KV) heads. \\
$N_g$ & Number of neuron groups in one MLP block. \\
$N_v$ & Number of neuron groups in the vocabulary of LM Head. \\
\bottomrule
\end{tabular}
\end{table}

\subsubsection{Problem Formulation}

Since the memory budget is the primary constraint determining inference feasibility, and the optimization goal is to minimize the heterogeneous cluster inference time, we formulate this as an integer linear programming (ILP) problem. The solution to this problem is the optimal workload schedule, which determines the number of neuron groups for each device. In summary, it can be formulated as an ILP problem with an objective function under constraint of each device's resources and LLM demands:

\small
\vspace{-10pt}
\begin{equation}
\begin{aligned}
& \underset{W}{\operatorname{argmin}} \quad \sum_{l=1}^{L} \sum_{k \in \{h, g\}} \max_{i=1,...,n}\left\{\frac{\tau_k \cdot W_{l,i,k}}{c_i}\right\} + 2L(n-1)T_{c} \\
&   \text{s.t.} \quad \sum_{i=1}^{n} W_{l,i,k} = N_k, \qquad \forall l \in \{1,...,L\}, \qquad k \in \{h, g\} \\
& \quad \quad \sum_{l=1}^{L} \sum_{k \in \{h, g\}} (W_{l,i,k} \cdot \text{mem}_k) \leq m_i, \qquad \forall i \in \{1,...,n\}. \\
\end{aligned}
\label{eq1}
\end{equation}

\normalsize
The two terms are computation and communication latency.
The LLM is composed of $L$ Transformer layers. Each layer contains $N_h$ attention heads in the MHA and $N_g$ neuron groups in the MLP. We aim to distribute this model across $n$ heterogeneous devices. 
The core of our optimization is the variable $W_{l,i,k}$, which is a non-negative integer representing the number of units of type $k$ (where $k \in \{h, g\}$) from layer $l$ that are assigned to device $i$. The computational cost to process a single attention head or a single MLP group is given by $\tau_h$ and $\tau_g$, respectively. Similarly, the memory required to store them is denoted by $\text{mem}_g$ and $\text{mem}_h$. In the formula, we use the generic terms $\tau_k$ and $\text{mem}_k$ to represent these costs depending on the component type $k$. 

For the communication latency, $2L(n-1)$ refers to the times of synchronization from participants to the master. $T_{c}$ represents the communication time required for synchronization between devices after each block's computation. Note that it is primarily determined by the hidden dimension $D$ and the link attributes, and thus is treated as a constant with respect to the workload distribution $W$. For instance, to calculate $T_{c}$, when $D=4096$ and link bandwidth is $1000Mbps$, the activation is $16KB$, then $T_{c}$ is $0.262ms$. The objective function in \Cref{eq1} is designed to minimize the overall latency.

\subsubsection{Semantic-Aware Predictors (SAPs)}\label{subsec:sparsity}
To reduce the communication overhead, we delve into the activation of neuron groups. 
Leveraging the insight on semantic imbalance and activation sparsity discussed in \Cref{insight1}, we design SAPs to predict critical neuron groups dynamically during runtime. Inspired by DEJAVU\cite{liu2023deja}, we gather the LLM activation and train our predictors.

Our SAPs differ from DEJAVU\cite{liu2023deja}'s predictors. 1). We use the activation norm as training data for fine-grained importance metric instead of processing it using sigmoid for two-class classification. 2). We adopt one layer ahead instead of the one block (MHA/MLP) ahead for overlap efficiency in \Cref{fig:predictor}. 3). We decrease the hidden dimension of predictors to 256 and 512 for 8B and 13B models instead of 1000. However, our modified predictors still have high performance; the SAPs training result achieves an average validation loss of 2.1\% and 2.4\% for MHA and MLP. Furthermore, we demonstrate the ability to maintain model accuracy in \Cref{sec:eval-acc}.

\subsubsection{Packet-Loss-Tolerant Fast Generation}\label{subsec:plr-generation}
This phase represents the actual inference stage. There are two designs to achieve end-to-end acceleration. Firstly, to eliminate the packet retransmission overhead, our relaxed synchronization design tolerates packet loss. The master node proceeds to merge available activations after a timeout period, rather than waiting for lost packets indefinitely and causing system deadlock. This timeout is set as $10ms$ empirically. 
By proactively assigning neuron groups predicted to be less important to devices with higher PLR, our method obviates costly retransmissions.

Second, to hide the latency of Pred and Load process, we design this overlap scheme as shown in \Cref{fig:predictor}.  
1). In the Pred process, each device's SAPs predict the next layer's importance rank of neuron groups locally using current semantic features as input. And then the importance ranks serve as a reference list for the neuron group priority indices, given in the \textit{Assignment} phase, to further determine mapping of local neuron group indices of the next block. 
2). In the Load and Comp process, each device will execute the assigned neuron groups and output its activation, which is two squares in the same color. Our method utilizes the linear layer's characteristics: Load time and Comp time both exhibit a linear relation with the number of neuron groups (i.e, parameter count). MLP has about 2 times the parameter count of MHA, which hinders the simplicity of the overlap scheme when the Load is one block ahead of Comp. In this case, the slower MLP Load will execute during the faster MHA Comp process. Similarly, the faster MHA Load will execute during the slower MLP Comp, causing a waste of waiting time. In contrast, when the Load is one layer ahead of the Comp, the latency of both processes for one block has a linear relation with the same number of neuron groups, thus they can be overlapped by easily allocating CPU threads for the two processes.
The limitation of this design is that each device should reserve a full KV cache locally in case needed by runtime-assigned neuron groups in MHA. But this memory overhead is minimal when the sequence length is not extremely long. For example, for Llama 8B model with sequence length 1024, KV cache requires $1024*L*2*D$ elements, only 3.3\% of the model size.
3). In the Comm process, for this example, the RPi has the highest PLR and is assigned the least important neuron groups. Our method schedules it to upload the least important activation, shown in the lightest color. As a result, the semantic features are not severely influenced even when the activation loss occurs. For the first layer, we use TCP because no semantic information can be given by SAPs. For the LM Head, we use TCP to ensure the logits are accurate.
Finally, the LLM's answer is a correct \texttt{\#3} in ~\Cref{fig:predictor}, showing that \Name delivers end-to-end acceleration while preserving accuracy.

\begin{algorithm}[t]
\caption{Computation-Greedy Scheduler}
\label{alg:comp-greedy}

\KwIn{$\boldsymbol{m} = \{m_1, \ldots, m_n\}$: Memory budgets of $n$ devices. $\boldsymbol{c} = \{c_1, \ldots, c_n\}$: Computing power of $n$ devices. $\mathcal{M}_t$: Model's total memory demand.}
\KwOut{$\boldsymbol{r}$: Memory allocation ratios vector.}

Assert {$\sum_{i=1}^n m_i \geq \mathcal{M}_t$}\;

$\boldsymbol{\sigma} \leftarrow \text{argsort}(\mathcal{C}, \text{direction='descending'})$ \;

$k \leftarrow \min\{j \mid \sum_{i=1}^j m_{\sigma_i} \geq \mathcal{M}_t\}$\; 
\Comment{Minimum number of devices needed}
$\mathcal{M}_k \leftarrow \mathcal{M}_t - \sum_{i=1}^{k-1} m_{\sigma_i}$ \;
\Comment{Memory needed for the last device}

$\boldsymbol{u}_i =
\begin{cases}
    m_i & \text{if } i \in \{\sigma_1, \ldots, \sigma_{k-1}\} \\
    \mathcal{M}_k & \text{if } i = \sigma_k \\
    0 & \text{otherwise}
\end{cases}
, \text{for } i=1, \ldots, n.$

$\boldsymbol{r} \leftarrow \{\boldsymbol{u}_i/\sum_{i=1}^n \boldsymbol{u}_i \mid i \in \{1,\ldots,n\}\}$ \;

\textbf{Return $\boldsymbol{r}$}\;
\end{algorithm}

\begin{algorithm}[th]
\caption{Min-Max Scheduler}
\label{alg:min-max}
\KwIn{$\boldsymbol{m}$. $\boldsymbol{c}$. $M_t$. $\epsilon$: Precision threshold ($10^{-3}$).}
\KwOut{$\boldsymbol{r}$: Memory allocation ratios vector.}

\SetKwFunction{BST}{SearchThreshold}
\SetKwFunction{MAR}{AllocateMemoryRatio}
\SetKwProg{Ft}{Function}{:}{}

\Ft{\BST{$\boldsymbol{m}, \boldsymbol{c}, M_t, \epsilon$}}{
    $left \leftarrow 0$, \quad $right \leftarrow {\max(\boldsymbol{m})}/{\min(\boldsymbol{c})}$ \;
    $\mathcal{T}_{\text{opt}} \leftarrow right$\; \Comment{Optimal utilization threshold}
    
    \While{$right - left > \epsilon$}{
        $mid \leftarrow {(left + right)}/{2}$ \;
        $Used\_Memory \leftarrow \sum_{i=1}^{n} \min(m_i, mid \cdot c_i)$ \;
        
        \uIf{$Used\_Memory \geq M_t$}{
            $\mathcal{T}_{\text{opt}} \leftarrow mid$, \quad $right \leftarrow mid$ \; \Comment{Found valid threshold}
        }
        \uElse{
            $left \leftarrow mid$; \Comment{Insufficient memory}
        }
    }
    \textbf{Return $\mathcal{T}_{\text{opt}}$}\; 
}
\Ft{\MAR{$\boldsymbol{m}, \boldsymbol{c}, \mathcal{T}$}}{
    $\boldsymbol{u} \leftarrow \{\min(m_i, \mathcal{T} \cdot c_i) \quad \text{ for } i \in \{1, 2, \ldots, n\}\}$; \Comment{ Calculate memory allocation}
    
    $\boldsymbol{r} \leftarrow \{({u}_i/\sum_{i=1}^n {u}_i) \quad \text{ for } i \in \{1, 2, \ldots, n\}\}$; \Comment{ Normalize to get ratios}
    
    \textbf{Return $\boldsymbol{r}$}\;
}
Assert {$\sum_{i=1}^n m_i \geq \mathcal{M}_t$}\;
$\mathcal{T}_{\text{opt}} \leftarrow$ \BST{$\boldsymbol{m}, \boldsymbol{c}, M_t, \epsilon$} \; 
\textbf{Return \MAR{$\boldsymbol{m}, \boldsymbol{c}, \mathcal{T}_{\text{opt}}$}}\;
\end{algorithm}

\subsubsection{PLR-Aware Load-balancing Scheduler (PLS)}\label{subsec:schedule}
PLS contains 2 steps to distribute workload by finding ratios and mapping indices. First, to solve the NP-hard ILP problem, we approximate it as a linear programming problem.
\Cref{alg:comp-greedy} is a greedy approach that first allocates tasks to the fastest devices until their full capacity is utilized, before assigning the remaining load to the next fastest devices in sequence. \Cref{alg:min-max} minimizes the maximum computation time among all devices to mitigate the straggler impact. It achieves this by maximizing the utilization of the entire parallel computing power. Second, \Cref{alg:scheduler} then maps the resulting workload ratios to neuron group priority indices, prioritizing network-stable devices for more critical activation.
For instance, to illustrate the function of \Cref{alg:scheduler}, consider partitioning 8 neuron groups among three devices, and the prioritized indices are $[1, ..., 8]$. Let the PLRs $\boldsymbol{\rho} = [0.0, 0.5, 0.1]$ and workload ratios $\boldsymbol{r} = [0.4, 0.4, 0.2]$. Then the result would be $[1, 2, 3], [6, 7, 8], [4, 5]$, which makes sure that the one with the highest PLR gets the least prioritized indices.

\begin{algorithm}[t]
\caption{PLR-Aware Ratio-Guided Scheduler}
\label{alg:scheduler}
\KwIn{Model dims \(N_h, N_{kv}, N_g, N_v\), Device count \(n\), ratios \(\boldsymbol{r}\), PLRs \(\boldsymbol{\rho}\).}
\KwOut{Scheduler \(\mathcal{S}\), which contains $n$ devices' partitioning plan for all neuron groups.}
\(\boldsymbol{q} \gets \text{argsort}(\boldsymbol{\rho}, \text{direction='ascending'})\) ; 

\(\boldsymbol{r}_{\text{sorted}} \gets \boldsymbol{r}[\boldsymbol{q}]\);  \Comment{Prioritized ratios}
\(\mathcal{S}_{\text{temp}} \gets \text{Partition}(\{N_h, N_{kv}, N_g, N_v\}, \boldsymbol{r}_{\text{sorted}})\)\;  \Comment{Partition the list of neuron groups indices according to $\boldsymbol{r}_{\text{sorted}}$}

\(\mathcal{S}[\boldsymbol{q}_k] \gets \mathcal{S}_{\text{temp}}[k] \quad \forall k \in \{1, \dots, n\}\)\;  \Comment{Map the k-th list of indices to the k-th prioritized device}
\textbf{Return} \(\mathcal{S}\)\;
\end{algorithm}

\section{Implementation}
\begin{figure}[t]
    \centering
    \includegraphics[width=\linewidth]{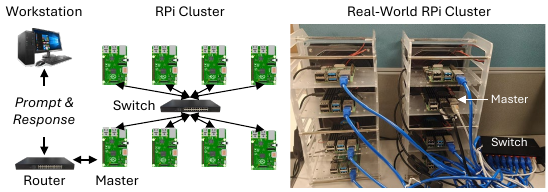}
    \caption{The Testbed Diagram and Real-World RPi Cluster.}
    \label{fig:testbed}
\end{figure}

We implemented \Name using 3.5k+ LOC in C++ atop \cite{dllama} and 1.4k+ LOC in Python atop \cite{gpt-fast}. The C++ implementation extends \texttt{dllama} (version 0.11.0)\cite{dllama}'s tensor parallelism with indexed model weight loading, hybrid communication backend, overlap functions, and timing statistics. The Python implementation modifies \cite{gpt-fast} with drop strategies and SAPs execution in the forward function and adds wrapper compatibility for the \texttt{lm-eval} framework. 
\Name is a standalone extensible framework and can be implemented using other DL frameworks such as TF-Lite, PyTorch. As shown in \Cref{fig:testbed}, the testbed consists of 8 RPis and one consumer-grade switch (Mercusys MS108G with Gigabit). One RPi is the master node responsible for the interactions with the workstation and coordinating the cluster. It also takes computing workload.

\section{Evaluation}\label{sec:eval}
\subsection{Experimental Setup}
\subsubsection{LLM and Dataset}
We evaluate our method on two open-source LLMs,  Llama-3.1-8B-Instruct and Llama-2-13B-Chat-Hf, both quantized to 4-bit integer precision. For brevity, their abbreviations are LlaMA3 and LlaMA2. Our goal is to accelerate model inference while maintaining task accuracy. 

We used three benchmark datasets to evaluate both prompt evaluation and text generation capabilities. ARC-Easy and ARC-Challenge \cite{clark2018think} are multiple-choice datasets, evaluated via the \texttt{loglikelihood} function. GSM8K test set \cite{cobbe2021gsm8k}, a challenging math task, requires reasoning (free-form answer generation) via the \texttt{generate\_until} function. 

\subsubsection{Edge Computing Scenario}

In the edge scenario, we evaluate the end-to-end LLM inference time. We formulate a subset of samples from the GSM8K \cite{cobbe2021gsm8k} dataset of sequence length 256. The prompt length is 128, and the generated token length is also 128. To emulate the network instability, we use the Linux tool \texttt{tc} to control the master node's link attribute. To emulate the heterogeneity, we adjust the CPU frequency to control the computing power precisely. \Cref{tab:rpi_specs} shows the specifications for the heterogeneity configurations. Each device CPU frequency ranges in 0.6-1.8 GHz with a step size of 0.1GHz; Each device memory budget ranges in 300-6300 MB with a step size of 200MB. These configurations are randomly generated under only two limitations: 1). the total memory budget is more than the LLM; 2). the average frequency is 1.2 GHz. For homogeneous configurations, all devices are identical in 1.8GHz frequency.

\begin{table}[t]
\centering
\caption{RPi Specifications and Configurations.}
\label{tab:rpi_specs}
\renewcommand{\arraystretch}{1.1} 
\begin{tabular}{|l|l|}
\hline
\textbf{Hardware} & \textbf{Specifications} \\
\hline 
CPU & 1.8 GHz Quad-core 64-bit SoC \\
\hline
RAM & 8GB LPDDR4 \\
\hline
\multirow{2}{*}{CPU Computing Power} & Range: 0.6-1.8 GHz; \\ 
& Step: 0.1 GHz \\
\hline
\multirow{2}{*}{Memory Budget} & Range: 300-6300 MB; \\ 
& Step: 200 MB \\
\hline
\end{tabular}
\renewcommand{\arraystretch}{1} 
\end{table}

\subsubsection{Baselines}
\textit{Vanilla TP} \cite{shoeybi2019megatron} is implemented by \cite{dllama} for all participated edge devices with evenly distribution.
\textit{Galaxy} \cite{ye2024galaxy} is a two-step scheduler that first allocates by computing power, then redistributes from OOM devices.

\subsubsection{Metrics}

\textit{Inference speed:} We measure the end-to-end generation time of LLM. Since there's little performance difference between prompt processing and decoding of the \texttt{dllama} \cite{dllama} framework, we gather the overall token count and inference time, then average as time per token (TPT). 
\textit{Model accuracy:} For evaluation, we employ the \texttt{lm-eval} library \cite{eval-harness} to evaluate model accuracy (ACC) on text tasks. Strict-match is adopted for answer correctness evaluation. We conduct experiments with four seeds to mitigate randomness and reported the average results.

\subsection{Accuracy Result}\label{sec:eval-acc}
\begin{table*}[t]
\setlength{\tabcolsep}{1mm}{}  %
\centering
\caption{Comparison of Model Accuracy across Different Models, Datasets and PLR Settings. The results unit is (ACC/\%) and the best results are bold. The upbound ACC of the dataset is in the parentheses after the model.
}
\resizebox{\linewidth}{!}{
\label{tab:main-acc-variation}
\begin{tabular}{l|c c c|c c c|c c c|c c c|c c c|c c c}
\toprule[1pt]
{Method} & \multicolumn{6}{c|}{ARC Easy} & \multicolumn{6}{c|}{ARC Challenge} & \multicolumn{6}{c}{GSM8K} \\
\midrule
\emph{Model} & \multicolumn{3}{c}{LLaMA2 (77.82\%)} & \multicolumn{3}{c|}{LLaMA3 (80.68\%)} & \multicolumn{3}{c}{LLaMA2 (46.08\%)} & \multicolumn{3}{c|}{LLaMA3 (50.17\%)} & \multicolumn{3}{c}{LLaMA2 (34.87\%)} & \multicolumn{3}{c}{LLaMA3 (68.16\%)} \\
\emph{PLR} & 0.01 & 0.02 & 0.05 & 0.01 & 0.02 & 0.05 & 0.01 & 0.02 & 0.05 & 0.01 & 0.02 & 0.05 & 0.01 & 0.02 & 0.05 & 0.01 & 0.02 & 0.05 \\
\midrule
Random & 75.46 & 74.33 & 73.23 & 77.69 & 75.93 & 75.08 & 45.73 & 44.37 & 44.11 & 48.04 & 48.46 & 45.73 & 33.21 & 33.43 & 31.24 & 62.17 & 67.85 & 64.82 \\

\Name & \textbf{77.36} & \textbf{77.40} & \textbf{76.94} & \textbf{80.26} & \textbf{80.05} & \textbf{79.12} & \textbf{46.25} & \textbf{45.99} & \textbf{45.99} & \textbf{50.85} & \textbf{50.60} & \textbf{50.85} & \textbf{35.33} & \textbf{35.10} & \textbf{34.57} & \textbf{68.92} & \textbf{68.54} & \textbf{68.01} \\

\bottomrule[1pt]
\end{tabular}
}
\end{table*}

\noindent\textbf{Stable performance across tasks and PLR settings.}
We evaluate ACC performance under varying PLR, where 25\% of model neuron groups may experience packet loss in \Cref{tab:main-acc-variation}. \Name maintains high ACC under different settings. Our method achieves state-of-the-art performance with an average 0.067\% ACC loss, compared with Random's 2.91\%.
Firstly, \Name performs excellently across different tasks, showing stable performance not only on knowledge-intensive tasks, ARC Easy and ARC Challenge, but also demonstrating clear advantages on mathematical reasoning problems GSM8K. This cross-task consistency indicates that \Name has good generalization capabilities and adaptability to out-of-domain application requirements. 
Second, \Name demonstrates more stable performance across different PLR settings, compared to the Random method. Notably, as the PLR increases (from 0.01 to 0.05), the accuracy of the Random method generally shows a downward trend, while our method maintains relatively stable performance. 
For example, on the ARC Easy dataset using the LLaMA2 model, the accuracy of the Random method drastically drops 4.59\% at PLR=0.05 (compared to 77.82\% upbound), while our method still maintains 76.94\%, slightly dropping 0.88\% from the upbound.

\noindent\textbf{Potential reasoning performance enhancement.} Our method demonstrates interesting performance improvement over upbound. Remarkably, for both LLaMA2 and LLaMA3, \Name at low PLR (0.01 and 0.02) surpasses the upbound on Arc Challenge and GSM8K. The improvement is significant for LLaMA3 on GSM8K, where \Name outperforms the upbound by 0.76\% at a 0.01 PLR. We speculate that this improvement stems from a regularization effect that mitigates overfitting. Analogous to Dropout, our method induces more streamlined and focused sub-networks by preserving only the most critical activations, eventually leading to enhanced accuracy.

\begin{figure}[t]
    \centering
    \includegraphics[width=0.85\linewidth]{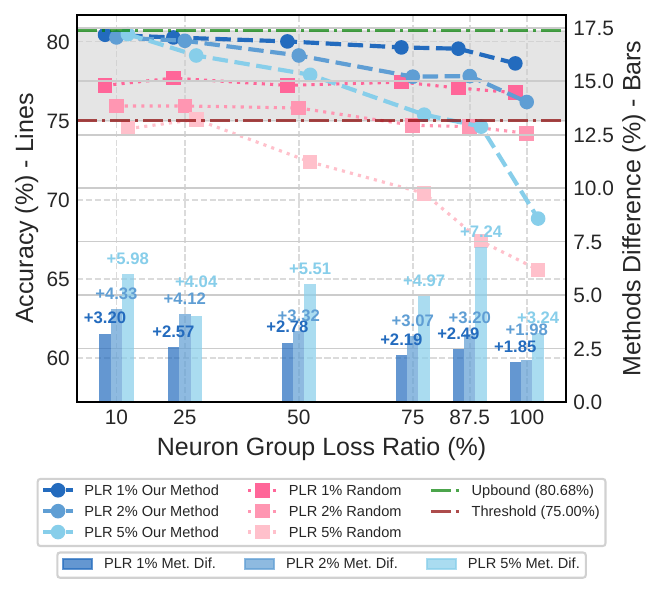}
    \caption{Comparison of LLaMA3 Accuracy on Arc Easy Dataset with \Name vs. Random Under Different PLR and Neuron Group Loss Ratios (NGLR).}
    \label{fig:model_robustness}
\end{figure}

\noindent\textbf{Robustness under extreme conditions.} 
To investigate the limit of our method's PLR tolerance, we report model performance variations under extreme loss scenarios, where PLR and NGLR gradually increase in this experiment \Cref{fig:model_robustness}. 
NGLR corresponds to $1-1/n$ when there are $n$ devices, $n-1$ of which experience packet loss. Scenarios of 100\% or less than 50\% NGLR is not practical because $n$ is a natural number but their theoretical accuracies are valuable.
As the lines show, \Name demonstrates remarkable robustness, maintaining an accuracy of around 75\%, equivalent to \textbf{93\%} of the unbound performance 80.68\%, across the vast majority of tested scenarios. Only one scenario leads to significant performance degradation: 100\% NGLR and 5\% PLR, indicating that one block (all neuron groups) could be skipped in 5\% probability. This scenario is not practical, as a distributed cluster would at minimum retain the activations from the master node, preventing a total 100\% loss. Therefore, our method ensures high accuracy in all plausible packet loss situations. In contrast, the Random method's performance begins to falter much earlier, dipping below 75\% accuracy with a PLR of just 2\% at higher NGLR. As shown in the bars of \Cref{fig:model_robustness}, most obviously at 5\% PLR and 87.5\% NGLR, \Name outperforms the Random method with the performance gap peaking at 7.24\%. This clearly showcases robustness and the superiority of \name.

\vspace{-6pt}
\subsection{Speed Result}\label{sec:eval-speed}
\begin{figure}
    \centering
    \includegraphics[width=0.7\linewidth]{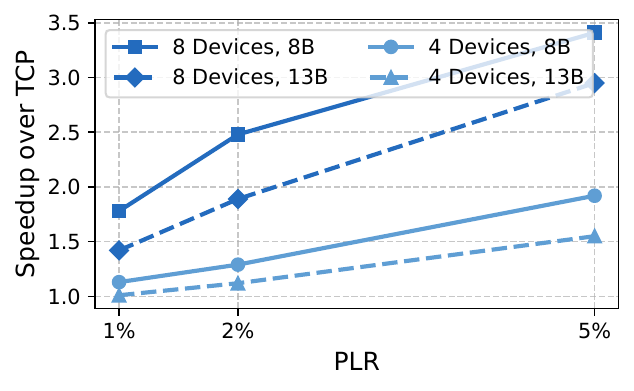}
    \caption{Speedup (\Name / TCP) of End-to-End Inference Time under Different Configurations. The legend refers to the number of participants and the model size.}
    \label{fig:TPT-comparison}
\end{figure}

\noindent\textbf{Communication acceleration under various settings.} \Cref{fig:TPT-comparison} demonstrates the speedup of end-to-end inference time (summation of computation time and communication time) across different configurations. Note that \Name only reduces communication time in this homogeneous experiment. The inference accuracy corresponds to the situation of 75\% and 87.5\% NGLR for 4 and 8 devices.
1). Our method consistently outperforms TCP across all configurations, with speedups ranging from 1.01× to 3.41×. As the PLR increases from 1\% to 5\%, the speedup becomes more pronounced, reaching up to \textbf{3.41×} with (8, 5\%) configuration, highlighting the scalability of our approach. 
2). A positive correlation exists between the number of participants and speedup. 8-device configurations, with higher communication time ratios, consistently yield superior acceleration compared to their 4-device counterparts.
3). 8B model obtains greater speedups compared to 13B model under the same settings. This is because larger models require higher computation time ratios, so the end-to-end speedup achieved by reducing communication time is less pronounced.

\begin{figure}[t]
    \centering
    \includegraphics[width=0.85\linewidth]{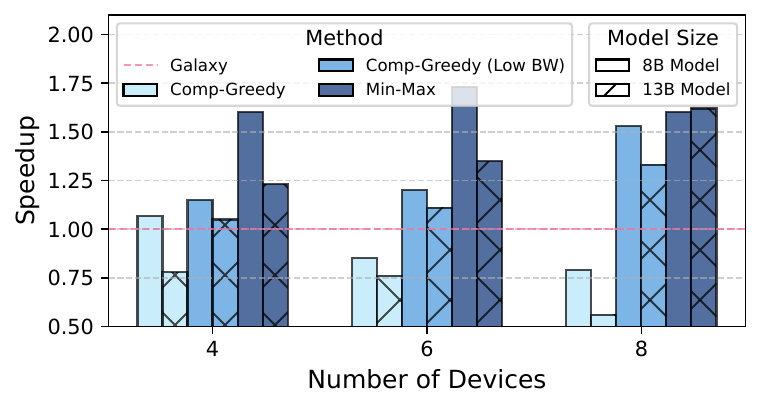}
    \caption{Average Speedup Across Multiple Heterogeneous Scenarios. Vanilla TP causes the OOM error and is not shown.}
    \label{fig:hetero-speedup}
\end{figure}

\noindent\textbf{Stable acceleration over heterogeneous devices.}
In \Cref{fig:hetero-speedup}, we evaluate $n$ random heterogeneous scenarios for each device count $n \in(4, 6, 8)$, and report the average speedups over Galaxy. The results demonstrate the consistent superiority of Min-Max across all scenarios, which achieves at most \textbf{1.73x} speedup when 6 devices run 8B model. 
For Comp-Greedy, it only performs better than Galaxy under low bandwidth (BW) 100Mbps situation by reducing the number of participants ($0$ in line 5 of \Cref{alg:comp-greedy}), hence decreasing $n$ in the communication term in \Cref{eq1}. Its worse performance under normal bandwidth originates from the lack of full utilization of slower devices. For Min-Max, its advantage becomes more pronounced, from 1.23x to 1.35x to 1.62x with 13B model, as the number of devices increases. This indicates that our ~\Cref{alg:min-max} (by finding $\mathcal{T}_{\text{opt}}$) becomes more effective with larger heterogeneity, particularly valuable for scaling. 

\subsection{System Overhead and Ablation}
This subsection investigates the overhead and/or function of each component of \name.

\paragraph{Overhead of SAPs and PLS}

1). SAPs Training Overhead. We utilize a small calibration set from SVAMP \cite{patel2021nlp}. The whole training time for a 13B model is 10min22s, negligible as a one-time operation. 
2). SAPs Inference Overhead. During inference, SAPs consistently occupies negligible memory and computation resources. For instance, the memory requirement for a quantized 8B model, SAPs's 37MB, is merely \textbf{0.82\%} of the model's 4.5GB when not considering the KV cache. This proportion is 0.32\% with a quantized 7.3GB 13B model.
3). Overhead of PLS.
Our PLS execution consumes about \textbf{11ms} on an RPi. This scheduling is performed as a one-off preprocessing task before each inference request. It is more efficient than solving the IIP problem.

\begin{table}[t]
    \centering
    \caption{Impact of Overlap Designs.}
    \begin{tabular}{l|c c|c}
        \toprule[1pt]
        \multirow{2}{*}{Overlap Design} & \multicolumn{2}{c|}{TPT Speedup} & \multirow{2}{*}{Average} \\
        & 8B & 13B & \\
        \midrule
        Baseline (No Overlap) & \multicolumn{3}{c}{1.00x} \\
        \midrule
        Pred-Comm Overlap Only & 1.10 & 1.25 & 1.18x \\
        Load-Comp Overlap Only & 1.14 & 1.11 & 1.13x \\
        Both Overlap Enabled & 1.22 & 1.38  & \textbf{1.30x} \\
        Upbound (No Pred\&Load) & 1.47 & 1.77 & 1.62x \\
        \bottomrule[1pt]
    \end{tabular}
    \label{tab:overlap-design-ablation}
\end{table}

\paragraph{Overlap Scheme Ablation}
In \Cref{tab:overlap-design-ablation}, we configure 3 threads for Comp/Pred and only 1 thread for Load/Comm. We conduct a detailed ablation study on the two overlap designs. The baseline is without any overlap. When implementing Pred-Comm overlap only, we observe a moderate speedup of 1.18x. Similarly, the Load-Comp overlap alone provides a speedup of 1.13x. The most significant performance gains are achieved by employing both overlap designs, delivering 1.22x speedup for the 8B model and 1.38x speedup for the 13B model. Notably, our combined overlap approach achieves \textbf{81.25\%} of the upper bound 1.62x. Due to the hardware limitation of RPi's quad-core, the configuration zoom is limited. But it could benefit from CPU with more threads and get better overlap efficiency. These results demonstrate that our overlap designs effectively minimize system-inherent overheads.

\section{Related Work}

\subsection{LLM on Single Device}
To tackle the challenge of massive memory footprint, LLM quantization can reduce the numerical precision of weights at the cost of accuracy. Frantar et al.\cite{frantar2022gptq} focuses on preserving accuracy through finding important weights and reducing their quantization degree. Shao et al.\cite{shao2023omniquant} aims to preserve activation outliers by learnable weight-activation quantization. Note that quantization methods are orthogonal to and compatible with \name.
Sparsity\cite{liu2023deja, liutraining} enables skipping less salient weights without compromising accuracy. Two studies \cite{alizadeh2024llm,song2024powerinfer} both adopt predictors from Liu et al.\cite{liu2023deja} to reduce memory footprint during inference by pre-fetching weights to be activated from external storage. However, these methods often do not suffice to accommodate large models inside a resource-constrained edge device. Sparsity can enhance robustness in our method's relaxed synchronization.

\subsection{Packet Loss Optimization}
In data centers, the TCP incast \cite{zhang2014modeling} is the main challenge for DNN training. Packet loss occurs in the congested switch buffer and motivates in-switch computing works\cite{lao2021atp,liu2024inart}. Specifically, Fei et al. propose \textit{OmniReduce}\cite{fei2021efficient}, similar to ours, by exploiting the gradient sparsity to eliminate the transmission overhead for zero-value packets. However, our method differs from theirs, which are solutions for DNN training and not directly applicable to edge LLM inference.
More fundamentally, the edge packet loss challenge stems not only from congestion but also from the unreliability of links\cite{ali2024congestion}. Lastly, our goal is to utilize sparsity for effective relaxed synchronization instead of reducing transmission packet count.

\section{Conclusion}
In this paper, we have proposed \name, a novel distributed LLM inference framework specifically tailored to unreliable edge networks.
From the semantic perspective, \Name assigns less critical neuron groups to unreliable edge nodes and introduces a parallel scheme to overlap the runtime prediction and the model loading processes. \Name adopts a PLR-aware scheduler incorporating load-balancing algorithms to optimize the utility of edge resources. 
Experiments on real-world testbed have demonstrated that \Name can maintain accuracy while accelerating the distributed LLM inference up to 3.41x in lossy edge network.

\bibliographystyle{IEEEtran}
\bibliography{ref}

@article{xu2024deploying,
  title={Deploying foundation model powered agent services: A survey},
  author={Xu, Wenchao and Chen, Jinyu and Zheng, Peirong and Yi, Xiaoquan and Tian, Tianyi and Zhu, Wenhui and Wan, Quan and Wang, Haozhao and Fan, Yunfeng and Su, Qinliang and others},
  journal={arXiv preprint arXiv:2412.13437},
  year={2024}
}

@article{king2024sasha,
  title={Sasha: creative goal-oriented reasoning in smart homes with large language models},
  author={King, Evan and Yu, Haoxiang and Lee, Sangsu and Julien, Christine},
  journal={Proceedings of the ACM on Interactive, Mobile, Wearable and Ubiquitous Technologies},
  volume={8},
  number={1},
  pages={1--38},
  year={2024},
  publisher={ACM New York, NY, USA}
}

@article{li2024personal,
  title={Personal llm agents: Insights and survey about the capability, efficiency and security},
  author={Li, Yuanchun and Wen, Hao and Wang, Weijun and Li, Xiangyu and Yuan, Yizhen and Liu, Guohong and Liu, Jiacheng and Xu, Wenxing and Wang, Xiang and Sun, Yi and others},
  journal={arXiv preprint arXiv:2401.05459},
  year={2024}
}

@inproceedings{patel2021nlp,
  title={Are NLP Models really able to Solve Simple Math Word Problems?},
  author={Patel, Arkil and Bhattamishra, Satwik and Goyal, Navin},
  booktitle={Proceedings of the 2021 Conference of the North American Chapter of the Association for Computational Linguistics: Human Language Technologies},
  pages={2080--2094},
  year={2021}
}

@article{li2024tpi,
  title={Tpi-llm: Serving 70b-scale LLMs efficiently on low-resource edge devices},
  author={Li, Zonghang and Feng, Wenjiao and Guizani, Mohsen and Yu, Hongfang},
  journal={arXiv preprint arXiv:2410.00531},
  year={2024}
}

@ARTICLE{9524600,
  author={Rischke, Justus and Sossalla, Peter and Itting, Sebastian and Fitzek, Frank H. P. and Reisslein, Martin},
  journal={IEEE Access}, 
  title={5G Campus Networks: A First Measurement Study}, 
  year={2021},
  volume={9},
  number={},
  pages={121786-121803},
  doi={10.1109/ACCESS.2021.3108423}}

@inproceedings{song2024powerinfer,
  title={Powerinfer: Fast large language model serving with a consumer-grade gpu},
  author={Song, Yixin and Mi, Zeyu and Xie, Haotong and Chen, Haibo},
  booktitle={Proceedings of the ACM SIGOPS 30th Symposium on Operating Systems Principles},
  pages={590--606},
  year={2024}
}

@inproceedings{alizadeh2024llm,
  title={Llm in a flash: Efficient large language model inference with limited memory},
  author={Alizadeh, Keivan and Mirzadeh, Seyed Iman and Belenko, Dmitry and Khatamifard, S and Cho, Minsik and Del Mundo, Carlo C and Rastegari, Mohammad and Farajtabar, Mehrdad},
  booktitle={Proceedings of the 62nd Annual Meeting of the Association for Computational Linguistics (Volume 1: Long Papers)},
  pages={12562--12584},
  year={2024}
}

@article{zhang2014modeling,
  title={Modeling and solving TCP incast problem in data center networks},
  author={Zhang, Jiao and Ren, Fengyuan and Tang, Li and Lin, Chuang},
  journal={IEEE Transactions on Parallel and Distributed systems},
  volume={26},
  number={2},
  pages={478--491},
  year={2014},
  publisher={IEEE}
}

@inproceedings{fei2021efficient,
  title={Efficient sparse collective communication and its application to accelerate distributed deep learning},
  author={Fei, Jiawei and Ho, Chen-Yu and Sahu, Atal N and Canini, Marco and Sapio, Amedeo},
  booktitle={Proceedings of the 2021 ACM SIGCOMM 2021 Conference},
  pages={676--691},
  year={2021}
}

@inproceedings{lao2021atp,
  title={ATP: In-network aggregation for multi-tenant learning},
  author={Lao, ChonLam and Le, Yanfang and Mahajan, Kshiteej and Chen, Yixi and Wu, Wenfei and Akella, Aditya and Swift, Michael},
  booktitle={18th USENIX Symposium on Networked Systems Design and Implementation (NSDI 21)},
  pages={741--761},
  year={2021}
}

@inproceedings{liu2024inart,
  title={InArt: In-Network Aggregation with Route Selection for Accelerating Distributed Training},
  author={Liu, Jiawei and Zhai, Yutong and Zhao, Gongming and Xu, Hongli and Fang, Jin and Zeng, Zhen and Zhu, Ying},
  booktitle={Proceedings of the ACM on Web Conference 2024},
  pages={2879--2889},
  year={2024}
}

@inproceedings{
frantar2022gptq,
title={{OPTQ}: Accurate Quantization for Generative Pre-trained Transformers},
author={Elias Frantar and Saleh Ashkboos and Torsten Hoefler and Dan Alistarh},
booktitle={The Eleventh International Conference on Learning Representations },
year={2023},
url={https://openreview.net/forum?id=tcbBPnfwxS}
}

@inproceedings{
shao2023omniquant,
title={OmniQuant: Omnidirectionally Calibrated Quantization for Large Language Models},
author={Wenqi Shao and Mengzhao Chen and Zhaoyang Zhang and Peng Xu and Lirui Zhao and Zhiqian Li and Kaipeng Zhang and Peng Gao and Yu Qiao and Ping Luo},
booktitle={The Twelfth International Conference on Learning Representations},
year={2024},
url={https://openreview.net/forum?id=8Wuvhh0LYW}
}

@misc{eval-harness,
  author       = {Gao, Leo and Tow, Jonathan and Abbasi, Baber and Biderman, Stella and Black, Sid and DiPofi, Anthony and Foster, Charles and Golding, Laurence and Hsu, Jeffrey and Le Noac'h, Alain and Li, Haonan and McDonell, Kyle and Muennighoff, Niklas and Ociepa, Chris and Phang, Jason and Reynolds, Laria and Schoelkopf, Hailey and Skowron, Aviya and Sutawika, Lintang and Tang, Eric and Thite, Anish and Wang, Ben and Wang, Kevin and Zou, Andy},
  title        = {A framework for few-shot language model evaluation},
  month        = 07,
  year         = 2024,
  publisher    = {Zenodo},
  version      = {v0.4.3},
  doi          = {10.5281/zenodo.12608602},
  url          = {https://zenodo.org/records/12608602}
}

@inproceedings{polo2024tinybenchmarks,
author = {Polo, Felipe Maia and Weber, Lucas and Choshen, Leshem and Sun, Yuekai and Xu, Gongjun and Yurochkin, Mikhail},
title = {tinyBenchmarks: evaluating LLMs with fewer examples},
year = {2024},
publisher = {JMLR.org},
booktitle = {Proceedings of the 41st International Conference on Machine Learning},
articleno = {1396},
numpages = {24},
location = {Vienna, Austria},
series = {ICML'24}
}

@article{cobbe2021gsm8k,
  title={Training Verifiers to Solve Math Word Problems},
  author={Cobbe, Karl and Kosaraju, Vineet and Bavarian, Mohammad and Chen, Mark and Jun, Heewoo and Kaiser, Lukasz and Plappert, Matthias and Tworek, Jerry and Hilton, Jacob and Nakano, Reiichiro and Hesse, Christopher and Schulman, John},
  journal={arXiv preprint arXiv:2110.14168},
  year={2021}
}

@misc{dllama,
  author = {Bartłomiej Tadych},
  title = {Distributed Llama},
  year = {2024},
  publisher = {GitHub},
  journal = {GitHub repository},
  howpublished = {\url{https://github.com/b4rtaz/distributed-llama}},
  commit = {7eb77ca93ec0d502e28d36b6fb20039b449cbea4}
}

@misc{gpt-fast,
  author = {{pytorch-labs}},
  title = {gpt-fast: Simple and efficient pytorch-native transformer text generation in <1000 LOC of python},
  year = {2024},
  publisher = {GitHub},
  journal = {GitHub repository},
  howpublished = {\url{https://github.com/pytorch-labs/gpt-fast}},
  commit = {[INSERT COMMIT HASH HERE]}
}

@inproceedings{ali2024congestion,
  title={Congestion or No Congestion: Packet Loss Identification and Prediction Using Machine Learning},
  author={Ali, Inayat and Hong, Seungwoo and Cheung, Taesik},
  booktitle={2024 International Conference on Platform Technology and Service (PlatCon)},
  pages={72--76},
  year={2024},
  organization={IEEE}
}

@inproceedings{bhadra2015packet,
  title={Packet loss probability in wireless networks: A survey},
  author={Bhadra, Dhwani R and Joshi, Charmi A and Soni, Priya R and Vyas, Nikita P and Jhaveri, Rutvij H},
  booktitle={2015 International Conference on Communications and Signal Processing (ICCSP)},
  pages={1348--1354},
  year={2015},
  organization={IEEE}
}

@inproceedings{sheshadri2017packet,
  title={On packet loss rates in modern 802.11 networks},
  author={Sheshadri, Ramanujan K and Koutsonikolas, Dimitrios},
  booktitle={IEEE INFOCOM 2017-IEEE Conference on Computer Communications},
  pages={1--9},
  year={2017},
  organization={IEEE}
}

@inproceedings{liu2023deja,
  title={Deja vu: Contextual sparsity for efficient llms at inference time},
  author={Liu, Zichang and Wang, Jue and Dao, Tri and Zhou, Tianyi and Yuan, Binhang and Song, Zhao and Shrivastava, Anshumali and Zhang, Ce and Tian, Yuandong and Re, Christopher and others},
  booktitle={International Conference on Machine Learning},
  pages={22137--22176},
  year={2023},
  organization={PMLR}
}

@inproceedings{liutraining,
  title={Training-Free Activation Sparsity in Large Language Models},
  author={Liu, James and Ponnusamy, Pragaash and Cai, Tianle and Guo, Han and Kim, Yoon and Athiwaratkun, Ben},
  booktitle={The Thirteenth International Conference on Learning Representations},
  year={2024},
}

@article{shoeybi2019megatron,
  title={Megatron-lm: Training multi-billion parameter language models using model parallelism},
  author={Shoeybi, Mohammad and Patwary, Mostofa and Puri, Raul and LeGresley, Patrick and Casper, Jared and Catanzaro, Bryan},
  journal={arXiv preprint arXiv:1909.08053},
  year={2019}
}

@article{zhang2024tinyllama,
  title={Tinyllama: An open-source small language model},
  author={Zhang, Peiyuan and Zeng, Guangtao and Wang, Tianduo and Lu, Wei},
  journal={arXiv preprint arXiv:2401.02385},
  year={2024}
}

@article{zhang2024edgeshard,
  title={Edgeshard: Efficient llm inference via collaborative edge computing},
  author={Zhang, Mingjin and Shen, Xiaoming and Cao, Jiannong and Cui, Zeyang and Jiang, Shan},
  journal={IEEE Internet of Things Journal},
  year={2024},
  publisher={IEEE}
}

@inproceedings{zhao2023lingualinked,
    title = "{L}ingua{L}inked: Distributed Large Language Model Inference on Mobile Devices",
    author = "Zhao, Junchen  and
      Song, Yurun  and
      Liu, Simeng  and
      Harris, Ian G.  and
      Abdu Jyothi, Sangeetha",
    editor = "Cao, Yixin  and
      Feng, Yang  and
      Xiong, Deyi",
    booktitle = "Proceedings of the 62nd Annual Meeting of the Association for Computational Linguistics (Volume 3: System Demonstrations)",
    month = aug,
    year = "2024",
    address = "Bangkok, Thailand",
    publisher = "Association for Computational Linguistics",
    url = "https://aclanthology.org/2024.acl-demos.16/",
    doi = "10.18653/v1/2024.acl-demos.16",
    pages = "160--171",
}

@inproceedings{ye2024galaxy,
  title={Galaxy: A Resource-Efficient Collaborative Edge AI System for In-situ Transformer Inference},
  author={Ye, Shengyuan and Du, Jiangsu and Zeng, Liekang and Ou, Wenzhong and Chu, Xiaowen and Lu, Yutong and Chen, Xu},
  booktitle={IEEE INFOCOM 2024-IEEE Conference on Computer Communications},
  year={2024},
  organization={IEEE}
}

@article{korthikanti2023reducing,
  title={Reducing activation recomputation in large transformer models},
  author={Korthikanti, Vijay Anand and Casper, Jared and Lym, Sangkug and McAfee, Lawrence and Andersch, Michael and Shoeybi, Mohammad and Catanzaro, Bryan},
  journal={Proceedings of Machine Learning and Systems},
  volume={5},
  pages={341--353},
  year={2023}
}

@article{clark2018think,
  title={Think you have solved question answering? try arc, the ai2 reasoning challenge},
  author={Clark, Peter and Cowhey, Isaac and Etzioni, Oren and Khot, Tushar and Sabharwal, Ashish and Schoenick, Carissa and Tafjord, Oyvind},
  journal={arXiv preprint arXiv:1803.05457},
  year={2018}
}
\end{document}